# On the recent seismic activity at Ioannina (Greece): Pre-earthquake electromagnetic emissions with critical and tricritical behavior

**Y. Contoyiannis[1], S. M. Potirakis[1], J. Kopanas[2], G. Antonopoulos[2], K. Eftaxias[3], C. Nomicos[4].**

1. Department of Electronics Engineering, Piraeus University of Applied Sciences (TEI of Piraeus), 250 Thivon & P. Ralli, GR-12244, Aigaleo, Athens, Greece, (Y.C.: yconto@uniwa.gr, S.M.P.: spoti@uniwa.gr)
2. Department of Environmental Technology and Ecology, Technological Education Institute (TEI) of the Ionian Islands, Panagoulas road, GR-29100, Zante, Greece, (J.K.: jkopan@otenet.gr, G.A.: sv8rx@teiion.gr)
3. Department of Physics, Section of Solid State Physics, University of Athens, Panepistimiopolis, GR-15784, Zografos, Athens, Greece, (ceftax@phys.uoa.gr)
4. Department of Electronics Engineering, Technological Education Institute (TEI) of Athens, Ag. Spyridonos, GR-12210, Aigaleo, Athens, Greece

## Abstract

In this short paper we report that the dynamics of the earthquake (EQ) preparation processes, as embedded in the observed fracture-induced electromagnetic emissions (EME), were characterized by critical and then tricritical behavior a few days before the occurrence of the recent Ioannina (Greece) earthquake [(39.79° N, 20.69° E), 15 October 2016, $M_L$=5.3]. Specifically, an excerpt of the MHz EME recorded by our remote telemetric station near the city of Ioannina five days before the EQ was found to possess critical characteristics, followed by an excerpt possessing tricritical characteristics. The analysis was performed by means of the method of critical fluctuations (MCF). Interestingly, spontaneous symmetry breaking was identified between critical and tricritical behavior.

## 1 Introduction

Earthquakes (EQs) are large-scale fracture phenomena in the Earth's heterogeneous crust. Fracture-induced electromagnetic emissions (EME) can be considered as the so-called precursors of general fracture. They are detectable both at a laboratory and a geophysical scale. These precursors allow a real-time monitoring of damage evolution during mechanical loading in a wide frequency spectrum ranging from kHz to MHz. Our main observational tool is the monitoring of the fractures which occur in the focal area before the final break-up by recording their kHz-MHz EME. Clear kHz-to-MHz EM anomalies have been detected over periods ranging from a few days to a few hours prior to recent destructive EQs in Greece and Italy. MHz EM anomalies has been found to be systematically emerging prior to kHz EM anomalies both in laboratory and geophysical scale (Eftaxias and Potirakis, 2013; Eftaxias et al., 2013, and references therein).

Based on the "Method of Critical Fluctuations" (MCF) (Contoyiannis and Diakonos, 2000; Contoyiannis et al., 2002, 2013), we have shown that the fracture-induced MHz EMEs recorded prior to recent significant EQs present criticality characteristics, implying that they emerge from a system in critical state (Contoyiannis et al., 2010; Potirakis et al., 2015, 2016).

The emerged "EM critical time window" can be described in analogy with a phase transition of second order in equilibrium. We note that using the natural time method, which can be successfully applied even if a limited number of data are available, we have shown that seismicity around the epicenter of the impending EQ presents also criticality characteristics (Potirakis et al., 2013, 2015, 2016). This finding indicates that these two different observables, which are measured in the same pre-seismic time period, namely, a few days before the EQ occurrence, could be attributed to the same Earth system, namely, the EQ generation process happening around the fault zone, enhancing the hypothesis for the possible relation of the recorded EME with the subsequent EQ. We have proposed that the observed MHz EM anomaly is due to the fracture of the highly heterogeneous system that surrounds the formation of strong brittle and high-strength entities (asperities) distributed along the rough surfaces of the main fault sustaining the system. We have also proposed that the finally emerged abrupt pulse-like kHz EM anomaly, which shows characteristics of a first order phase transition is due to the fracture of the aforementioned large high-strength entities themselves (Contoyiannis et al, 2015).

In statistical physics, a tricritical point is a point in the phase diagram of a system at which the two basic kinds of phase transition, that is the second order transition and the first order transition. A characteristic property of the area around this point is the co-existence of three phases, specifically, the symmetry area, the low symmetry phase, and an intermediate "mixing" phase. A passage through the area, around the tricritical crossover, from the second order phase transition to the first order transition through the intermediate "mixing" state constitutes a tricritical crossover. Thus, in terms of fracture-induced EME, such a situation would be manifested by an "EM tricritical time window" emerging in the observed preseismic EME time series after the appearance of an "EM critical time window". The appearance of the aforementioned EME tricritical signature indicates that the progress to the first order phase transition of the fracture system, namely, the EQ occurrence, is approaching (Contoyiannis et al, 2015; Potirakis et al., 2016).

Herein, we report the recording of a MHz EME signal prior to a recent significant EQ occurred in northeast Greece, near the city Ioannina. More precisely, on 15 October, 2016 (20:14:49 UT), an $M_L$=5.3, shallow (17km), EQ took place near the aforementioned city (39.79$^o$ N, 20.69$^o$ E), while 5 days before, the Ioannina station (cf. Fig. 1) of our telemetric network recorded MHz EME signals. This detection gives a good reason to examine whether the above mentioned scenario of the EQ preparation process in terms of preseismic EME is also verified in the case of the Ioannina EQ. Our analyses show that this indeed happens, a "critical time window" is included in the detected MHz EM anomaly which has been followed by a "tricritical time window". Unfortunately, at the time, all neighboring stations of our telemetric network (cf. Fig. 1), i.e., Kozani, Kerkira (Corfu), Kefalonia (Cephalonia) and Zakynthos (Zante), were out of order. So, no more EME signals were available for similar analysis.

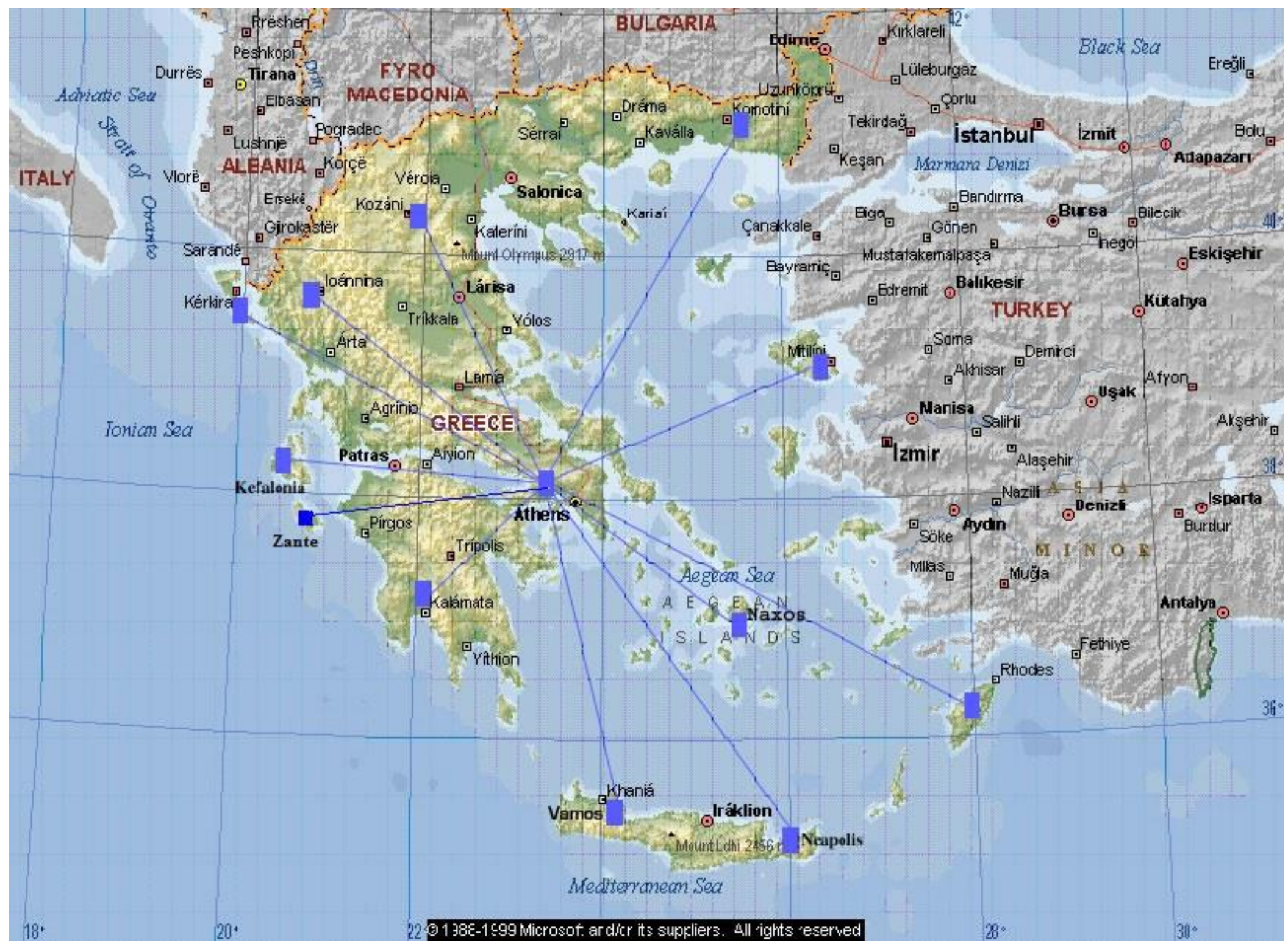

**Fig. 1.** The 11 remote sensing stations completing the telemetric network for the recording of electromagnetic variations in the MHz and kHz bands in Greece [details for the network and the involved instrumentation are available in the online supplementary downloadable material of (Potirakis et al., 2013)].

## 2 Data analysis method

The analysis of the recorded data was performed using the method of critical fluctuations (MCF) (Contoyiannis and Diakonos, 2000; Contoyiannis et al., 2002, 2013). Detailed descriptions of all the involved calculations can be found elsewhere (Contoyiannis et al., 2013) and therefore are omitted here for the sake of brevity and focus on the findings. However, a general description of the employed method follows.

MCF was proposed for the analysis of critical fluctuations in the observables of systems that undergo a continuous phase transition (Contoyiannis and Diakonos, 2000; Contoyiannis et al., 2002). It is based on the finding that the fluctuations of the order parameter, that characterizes successive configurations of critical systems at equilibrium, obey a dynamical law of intermittency of an 1D nonlinear map form. The MCF is applied to stationary time windows of statistically adequate length, for which the distribution of the of waiting times $l$ (laminar lengths) of fluctuations in a properly defined laminar region is fitted by a function $f(l) \propto l^{-p_2} e^{-p_3 l}$. The criteria for criticality are $p_2 > 1$ and $p_3 \approx 0$ (Contoyiannis and Diakonos, 2000; Contoyiannis et al., 2002), while for tricriticality are $p_2 < 1$ and $p_3 \approx 0$ (Contoyiannis et al, 2015).

## 3 Analysis results

Part of the MHz recordings of the Ioannina station associated with the recent $M_L = 5.3$ EQ is shown in Fig. 2a. This was recorded in Julian day 284, that is ~5 days before the occurrence of the recent Ioannina EQ. This stationary time-series excerpt, having a total length of ~5.6h (20000 samples) starting at 10 Oct. 2016 (00:00:01 UT), was analyzed by the MCF method and was identified to be a "critical window" (CW). CWs are time intervals of the MHz EME signals presenting features analogous to the critical point of a second order phase transition (e.g., Contoyiannis et al, 2015).

The main steps of the MCF analysis (e.g., Contoyiannis et al, 2015; Potirakis et al., 2016) on the specific time-series are shown in Fig. 2b- Fig. 2d. As Fig. 2d shows the obtained plot of the $p_2$, $p_3$ exponents vs. $\phi_l$ apparently satisfy the criticality conditions, $p_2 > 1$ and $p_3 \approx 0$, for a wide range of end points $\phi_l$, revealing the power-law decay feature of the time-series that proves that the system is characterized by intermittent dynamics; in other words, the MHz time-series excerpt of Fig. 2a is indeed a CW.

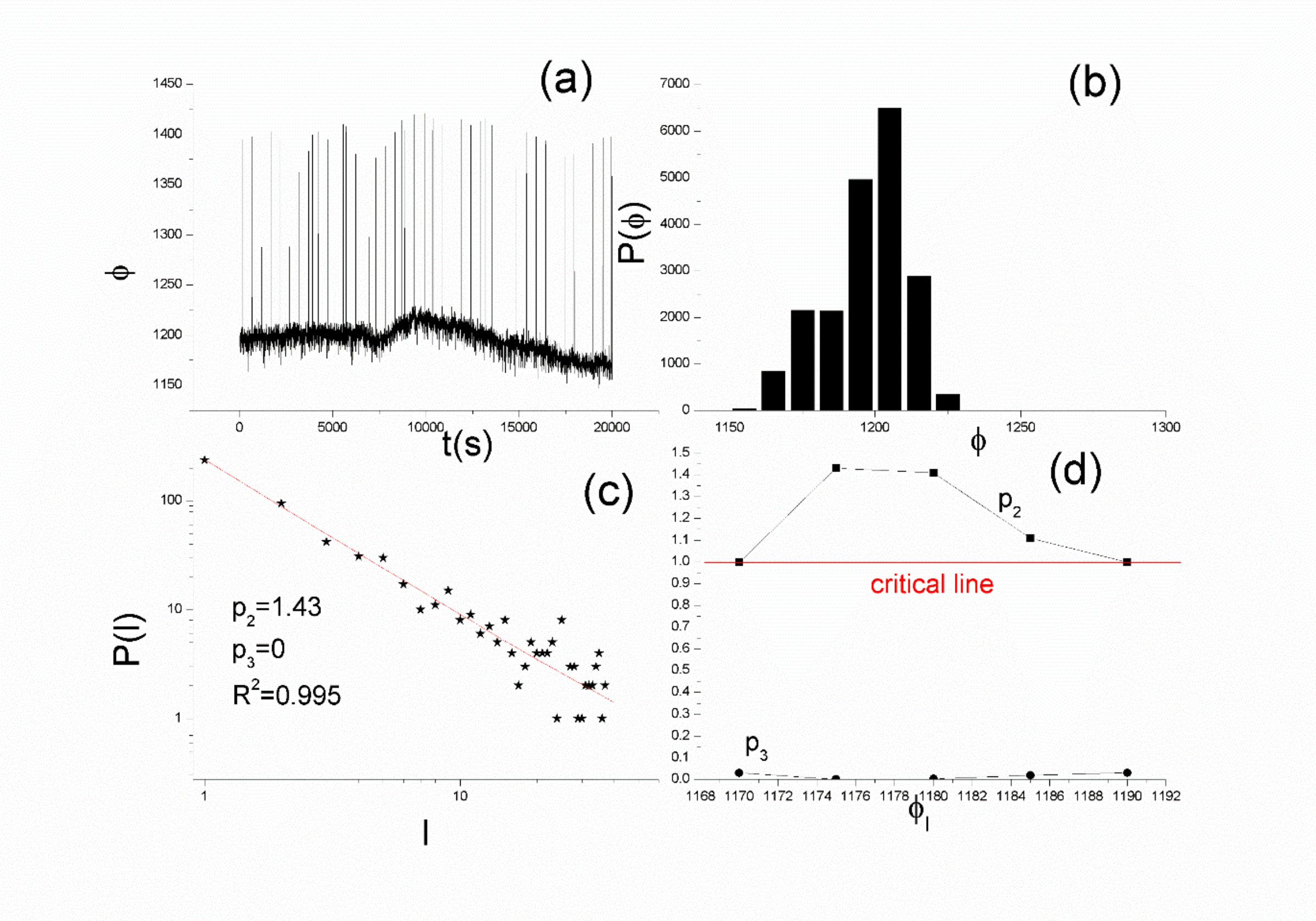

**Fig. 2**. (a) The 20000 samples long critical window of the MHz EME that was recorded before the Ioannina EQ. (b) Amplitude distribution of the signal of Fig. 2a. (c) Representative example of the involved fitting for the laminar distribution resulting for a specific end point. The solid line corresponds to the fitted function (cf.

to text in Sec. 2) with the values of the corresponding exponents $p_2$, $p_3$ also noted. (d) The obtained exponents $p_2$, $p_3$ vs. different values of the end of laminar region $\phi_l$. The horizontal dashed line indicates the critical limit ($p_2 = 1$).

A few hours after the emergence of the CW, part of the MHz recordings of the same station presented tricritical characteristics, as shown by Fig. 3 where the tricriticality conditions $p_2 < 1$ and $p_3 \approx 0$ are clearly satisfied for a wide range of laminar lengths. This EME time series excerpt has a total length of ~6.1h (22000 samples) starting at 10 Oct. 2016 (16:06:40 UT).

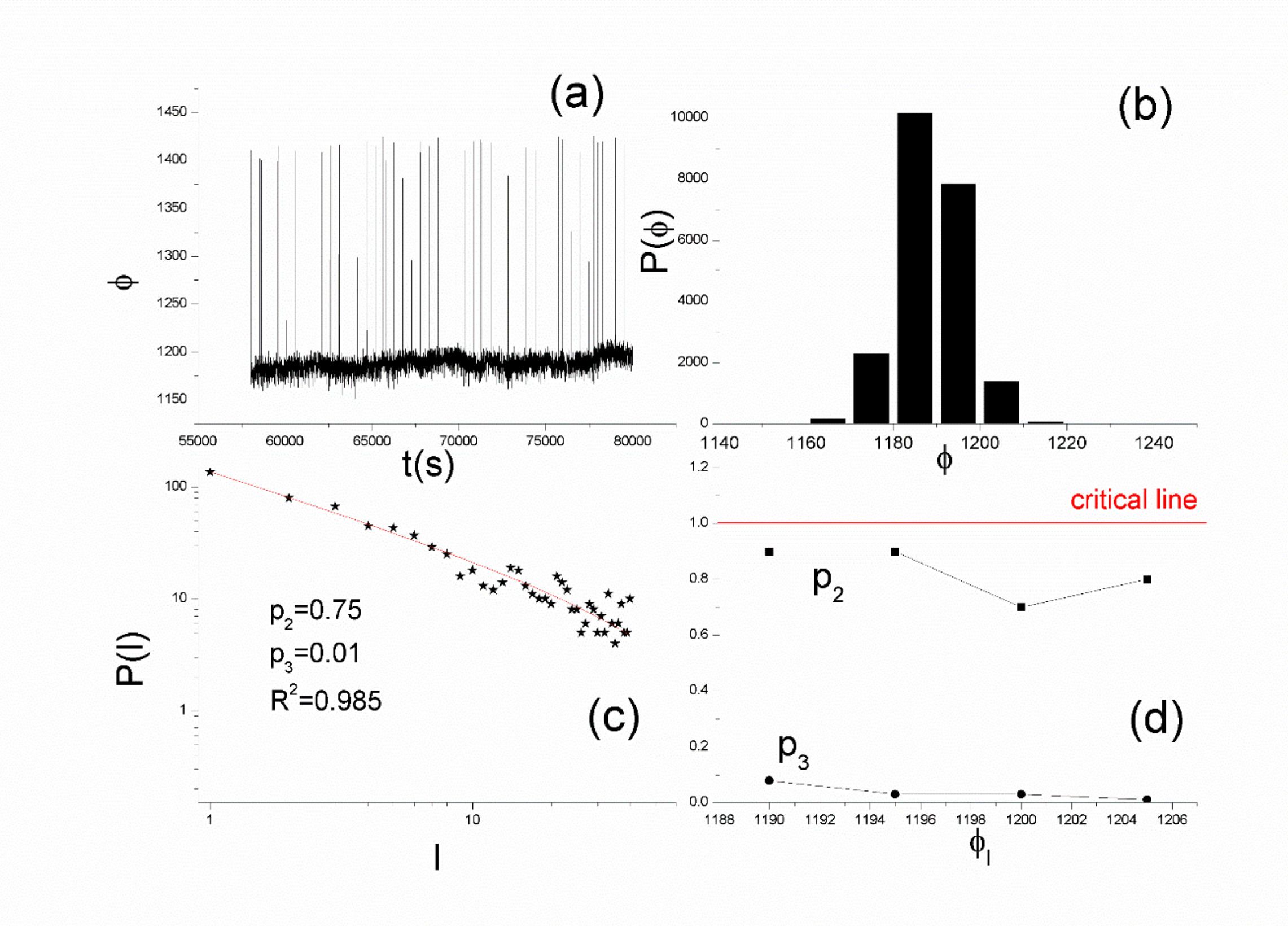

**Fig. 3**. (a) The 22000 samples long tricritical window of the MHz EME that was recorded prior to the Ioannina EQ. (b) Amplitude distribution of the signal of Fig. 3a. (c) Representative example of the involved fitting for the laminar distribution resulting for a specific end point. The solid line corresponds to the fitted function (cf. to text in Sec. 2) with the values of the corresponding exponents $p_2$, $p_3$ also noted. (d) The obtained exponents $p_2$, $p_3$ vs. different values of the end of laminar region $\phi_l$. The horizontal dashed line indicates the critical limit ($p_2 = 1$).

A new finding resulting from the analysis of the specific preseismic MHz EME is that the tricritical window appears after spontaneous symmetry breaking of a second-order phase transition (Figs. 4 and 5). Identifying a tricritical window in a MHz EME is relatively rare. However, we have to note that, as it has already been observed for a number of cases (not yet published), when a tricritical windows is identified in MHz EME it is preceded by symmetry breaking (even symmetry breakings). This is an interesting property that needs to be further investigated with the help of thermal systems undergoing a second-order phase transition.

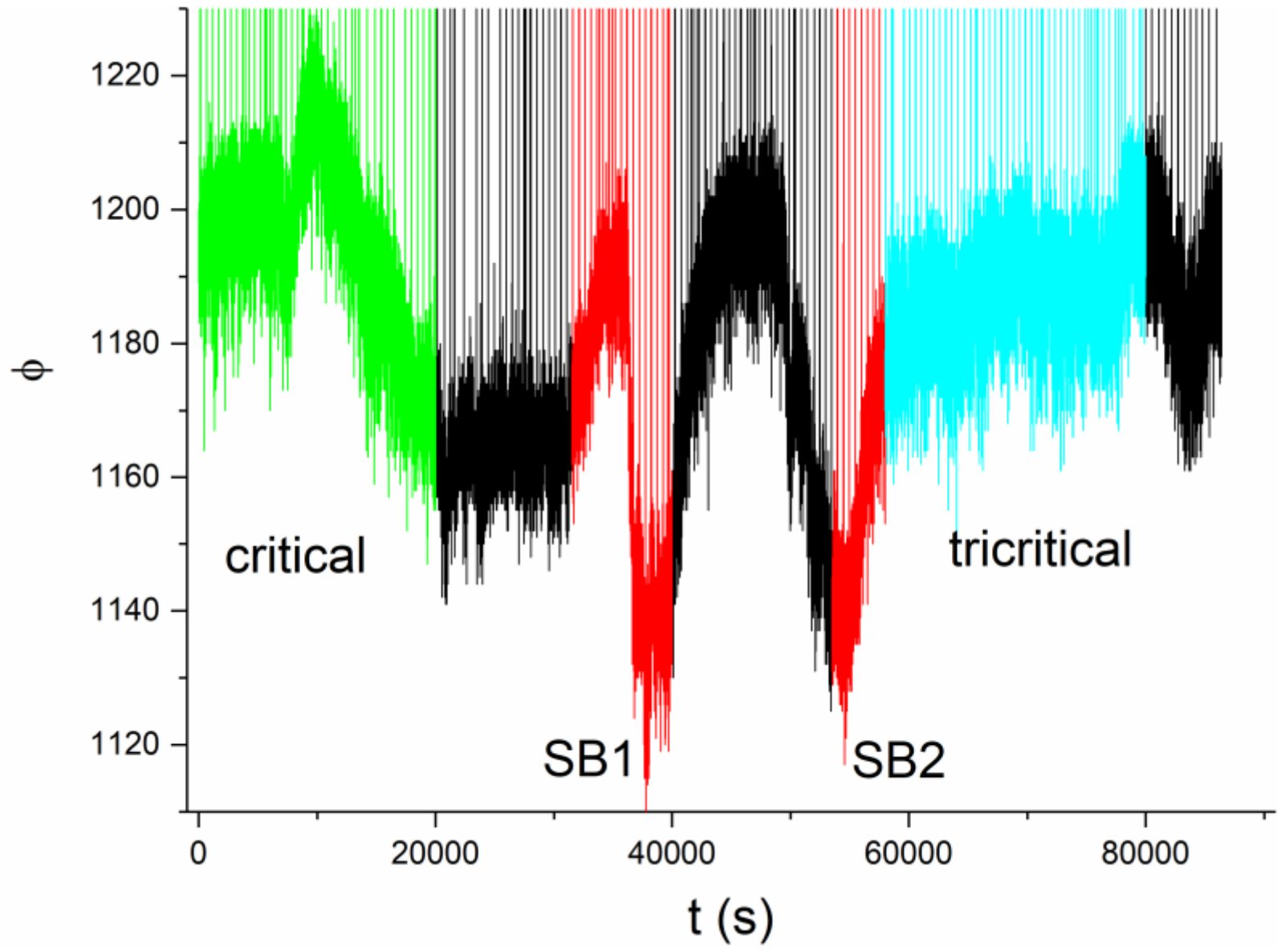

**Fig. 4**. The critical window, the analysis of which has been presented in the Fig. 2, is shown in green. The tricritical window, the analysis of which has been presented in the Fig. 3, is shown in blue. The two time series excerpts corresponding to spontaneous symmetry breaking are shown in red. The time sequence is the following: the critical window is followed by symmetry breaking of the second-order phase transition, and immediately after that the tricritical window appears.

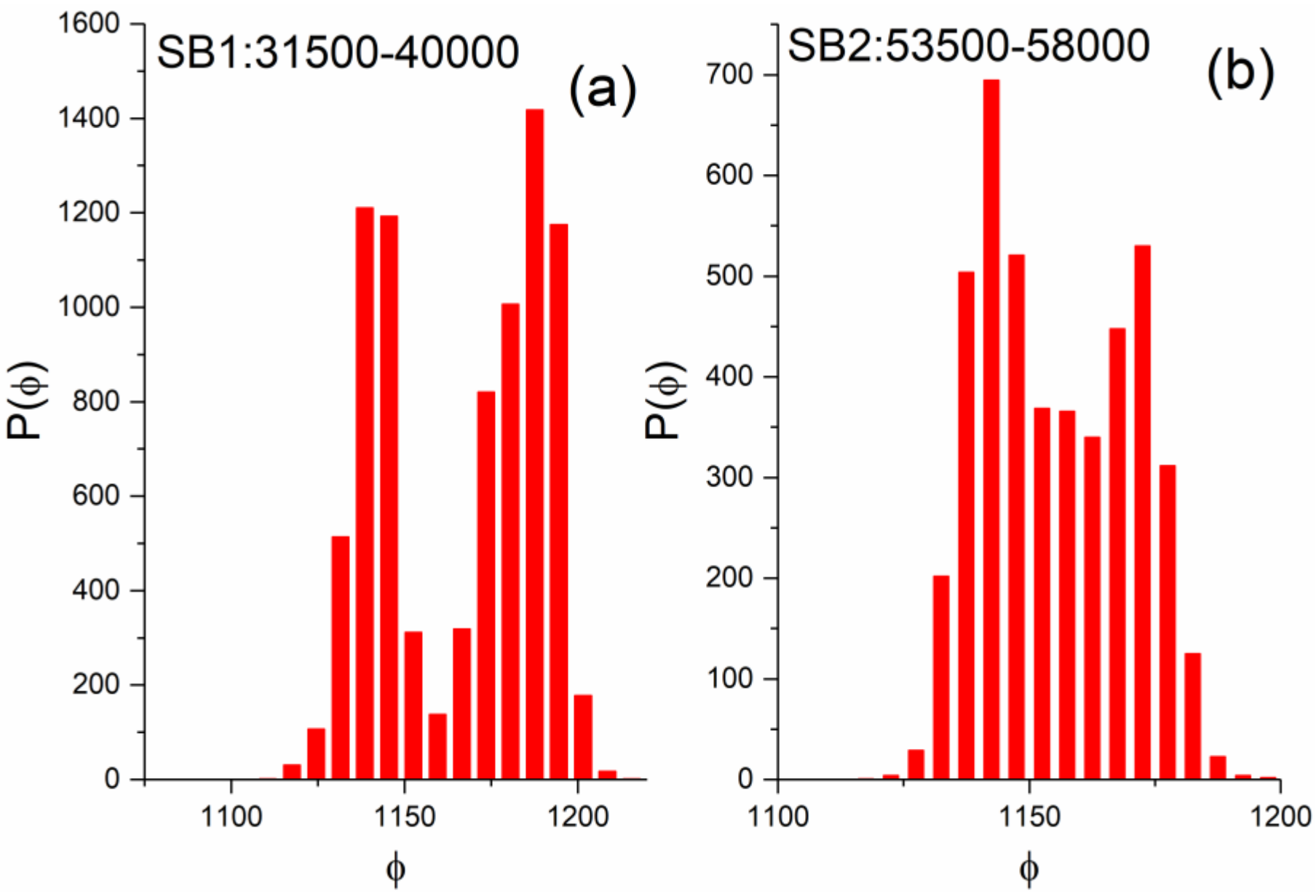

**Fig. 5**. (a) The two lobes resulting from the spontaneous symmetry breaking of a second-order phase transition for finite systems have almost completely separated. (b) The two lobes resulting from the spontaneous symmetry breaking of a second-order phase transition for finite systems communicate with each other. After this symmetry breaking process, the tricritical window follows.

## 4 Discussion - Conclusions

Based on the method of critical fluctuations, we have shown that the fracture-induced MHz EME recorded prior to the recent significant EQ of Ioannina successively present criticality and trictiticality characteristics. The first observed critical behavior of the EME time series implies that this has been generated by a system in critical state. The next appeared triciritical MHz EME indicates that the progress to the first order phase transition of the fracture system, namely, the EQ occurrence, is approaching. Notably, the triciritical MHz EME emerges after indications of spontaneous symmetry breaking of a second-order phase transition. This is an interesting property that needs to be further investigated with the help of thermal systems undergoing a second-order phase transition.